\documentclass[fleqn,10pt]{wlscirep}

\usepackage{ulem}
\usepackage{sidecap}
\title{CeRu$_4$Sn$_6$: a strongly correlated material with nontrivial topology}

\author[1]{Martin~Sundermann}
\author[1]{Fabio~Strigari}
\author[1]{Thomas~Willers}
\author[2]{Hannes~Winkler}
\author[2]{Andrey~Prokofiev}
\author[3]{James~M.~Ablett}
\author[3]{Jean-Pascal~Rueff}
\author[4]{Detlef~Schmitz}
\author[4]{Eugen~Weschke}
\author[5]{Marco~Moretti~Sala}
\author[5]{Ali~Al-Zein}
\author[6]{Arata~Tanaka}
\author[7]{Maurits~W.~Haverkort}
\author[7]{Deepa~Kasinathan}
\author[7]{Liu Hao~Tjeng}
\author[2]{Silke~Paschen}
\author[1,*]{Andrea~Severing}

\affil[1]{Institute of Physics II, University of Cologne, Z{\"u}lpicher Stra{\ss}e 77, 50937 Cologne, Germany}
\affil[2]{Institute of Solid State Physics, Vienna University of Technology, Wiedner Hauptstr. 8-10, 1040 Vienna, Austria}
\affil[3]{Synchrotron SOLEIL, L'Orme des Merisiers, Saint-Aubin, BP 48, 91192 Gif-sur-Yvette C\'edex, France}
\affil[4]{Helmholtz-Zentrum Berlin, BESSY II, Albert-Einstein-Stra{\ss}e 15, 12489 Berlin, Germany}
\affil[5]{European Synchrotron Radiation Facility (ESRF), B.P. 220, 38043 Grenoble C\'edex, France}
\affil[6]{Department of Quantum Matter, ADSM Hiroshima University, Higashi-Hiroshima 739-8530, Japan}
\affil[7]{Max Planck Institute for Chemical Physics of Solids, N{\"o}thnizer Stra{\ss}e 40, 01187 Dresden, Germany}
\affil[*]{severing@ph2.uni-koeln.de}

\begin{abstract}
Topological insulators form a novel state of matter that provides new opportunities to create unique quantum phenomena. While the materials used so far are based on semiconductors, recent theoretical studies predict that also strongly correlated systems can show non-trivial topological properties, thereby allowing even the emergence of surface phenomena that are not possible with topological band insulators. From a practical point of view, it is also expected that strong correlations will reduce the disturbing impact of defects or impurities, and at the same increase the Fermi velocities of the topological surface states. The challenge is now to discover such correlated materials. Here, using advanced x-ray spectroscopies in combination with band structure calculations, we infer that CeRu$_4$Sn$_6$ is a strongly correlated material with non-trivial topology. 
\end{abstract}
\begin{document}

\flushbottom
\maketitle

\thispagestyle{empty}


\section*{Introduction}

A metallic surface is one of the necessary characteristics of a 3-dimensional topological insulator. The surface charge carriers have a massless Dirac dispersion and are protected from backscattering by time reversal symmetry \cite{Kane_2005,Fu_2007,Zhang_2009}. So far the materials used in the research field of topological insulators are based exclusively on simple semiconductors and only very recently it was realized that also strongly correlated systems can be topologically non-trivial. Here the Kondo or intermediate-valent insulator SmB$_6$\cite{Menth_1969} came immediately into consideration\cite{Dzero_2010, Takimoto_2011, Dzero_2012, Dzero_2013, Alexandrov_2013, Lu_2013} and first investigations\cite{Jiang_2013,Frantzeskakis_2013,Kim_2014,Li_2014,Xu_2014,Ishida_2015} show that the robust metallicity of the sample surface could be a promising concept for explaining its long-time mysterious low temperature residual conductivity\cite{Riseborough2000}. Increasing the interest for correlated topological insulators is the possibility to find phenomena that might be even more exotic\cite{PhysRevX.3.011016,1742-5468-2013-09-P09016,PhysRevB.88.115137,PhysRevX.3.041016} than the ones of topological semiconductors. 

The important issue is now to find other strongly correlated materials with non-trivial topology. This is not a trivial task since quantitative numerical tools often applied to characterize the topological nature of semiconductors are lacking for correlated materials, due to the fact that the reliability and accuracy of mean field solutions of the electronic structure in the presence of strong correlations are far from sufficient.
It is therefore important that recent theoretical studies have 
drafted effective models for strongly correlated systems that should produce topological non-trivial behavior\cite{Dzero_2010, Takimoto_2011,Lu_2013}. One of the propositions is that rare earth Kondo or mixed-valent insulators should have all the necessary ingredients for realizing topological insulating properties\cite{Dzero_2010, Takimoto_2011, Dzero_2012, Dzero_2013, Alexandrov_2013, Lu_2013}: a~fully spin-compensated insulating ground state formed by orbitals having opposite parity, e.g. $4f$ vs. $5d$, and inverted in energy in particular sections in the Brillouin zone. It is even specifically mentioned that for Ce systems with cubic\cite{Dzero_2010,  Takimoto_2011} and tetragonal symmetry\cite{Dzero_2012, Dzero_2013} a gap always opens when the ground state is a nodeless $J_z$\,=\,$\pm 1/2$ Kramers doublet and when the $4f$ occupation is sufficiently away from integer.

Here CeRu$_4$Sn$_6$ comes into play. It is a tetragonal, non-centrosymmetric ($I\bar{4}2m$) \cite{Pottgen_1997}, potential Kondo insulator with a $J$\,=\,5/2 Hund's rule ground state. A hybridization-induced gap was reported first in 1992\,\cite{Das_1992}, supported further by several experimental studies\,\cite{Strydom_2005,Paschen_2010,Bruning_2010}, and later estimated to be of the order of 100\,K from single crystal resistivity data\,\cite{Paschen_2010,Winkler_2012}. Optical conductivity measurements point towards remnant metallicity\,\cite{Guritanu_2013} and yet the low temperature resistivity is in the $milli$Ohm-cm range, indicating clearly the depletion of the charge carriers\,\cite{Das_1992,Paschen_2010,Winkler_2012,Guritanu_2013}. Magnetization and resistivity are highly anisotropic strongly suggesting the $J$\,=\,5/2 Hund's rule ground state is split by the tetragonal crystal-electric field (CEF)\,\cite{Paschen_2010,Winkler_2012}. $\mu$SR measurements did not find magnetic order down to 50\,mK, thus pointing out the importance of Kondo-type interactions in this material \,\cite{Strydom_2007} .

One of the pressing open questions is whether CeRu$_4$Sn$_6$ fulfills the conditions mentioned in the theoretical studies about strongly correlated topological insulators\,\cite{Dzero_2010, Takimoto_2011, Dzero_2012, Dzero_2013, Alexandrov_2013, Lu_2013}. To answer this question, we determine here whether the hybridization between conduction and $4f$ electrons ($cf$-hybridization) in CeRu$_4$Sn$_6$ is strong enough and whether the CEF ground state has a favorable symmetry to develop a gap of sizable magnitude. The degree of hybridization has been investigated by performing a temperature dependent x-ray absorption spectroscopy (XAS) experiment at the Ce $L$-edge to obtain the Ce valence and the Kondo temperature $T_K$. The CEF ground state symmetry was studied by measuring the Ce $M$-edge in an XAS experiment and the Ce N-edge in a non-resonant inelastic x-ray scattering (NIXS) experiment. We have also performed band structure calculations to understand better our experimental findings and to test the robustness of our conclusions. Details about the experimental set-ups and the single crystal preparation and quality can be found in the Section \textsl{Methods}.

\section*{Results}
\subsection*{Cerium $4f$ valence and Kondo temperature}
Trivalent cerium (Ce$^{3+}$) has a $4f$ occupancy $n_f$\,=\,1 ($f^1$). The presence of strong $cf$-hybridization leads to a partial delocalization of the $4f$ electrons into the conduction electron sea and the resulting non-integer valent $4f$ ground state can be written in a local picture as a mixed state $|\Psi_{\mathrm{GS}}\rangle$\,=\,$c_0$\,$|f^0\rangle$\,+\,$c_1$\,$|f^1$\uline{L}$\rangle$\,+\,$c_2$\,$|f^2$\uuline{L}$\rangle$, i.e$.$ with additional contributions of tetravalent ($f^0$) and divalent ($f^2$) states\cite{GunnarssonPRB28}. Here \uline{L} and \uuline{L} denote the number of ligand holes. For Ce the resulting occupation of the $4f$ shell is then given by $n_f$\,=\,2$c_2^2$\,+\,$c_1^2$\,$\leq$\,1, whereas $c_0^2$ quantifies the degree of delocalization. The hybridization is described in terms of the formation of a Kondo singlet state at $k_B T_K$ below the energy of the $4f^1$ configuration, which breaks down as thermal fluctuations match the same energy. Hence, information about the Kondo temperature $T_K$ is obtained from the temperature variation of $n_f$, or equivalently of $c_0^2$ if $c_2^2$\,$\ll$\,$c_1^2$ or if $c_2^2$ does not change with $T$\cite{Bickers_1987,Tjeng_1993,Dallera_2002}.

Core level spectroscopy is a powerful tool for determining the occupations of the various $4f^n$ configurations\cite{GunnarssonPRB28}. This is based on the fact that the impact of the core-hole potential depends strongly on the configuration. Moreover, recent XAS experiments measuring the $L_{\alpha_1}$ recombination process ($3d_{5/2} \rightarrow 2p_{3/2}$) with partial fluorescence yield (PFY-XAS) at the Ce L$_3$-edge ($2p \rightarrow 5d$) have tremendously improved the accuracy of determining valencies\cite{Kotani_2001,Dallera_2002,Rueff_2010}. We have performed such a PFY-XAS experiment on CeRu$_4$Sn$_6$, determined $c_0^2$ and obtained $T_K$ from its temperature dependence.

\begin{SCfigure}
	\centering
	\includegraphics[width=0.5\linewidth]{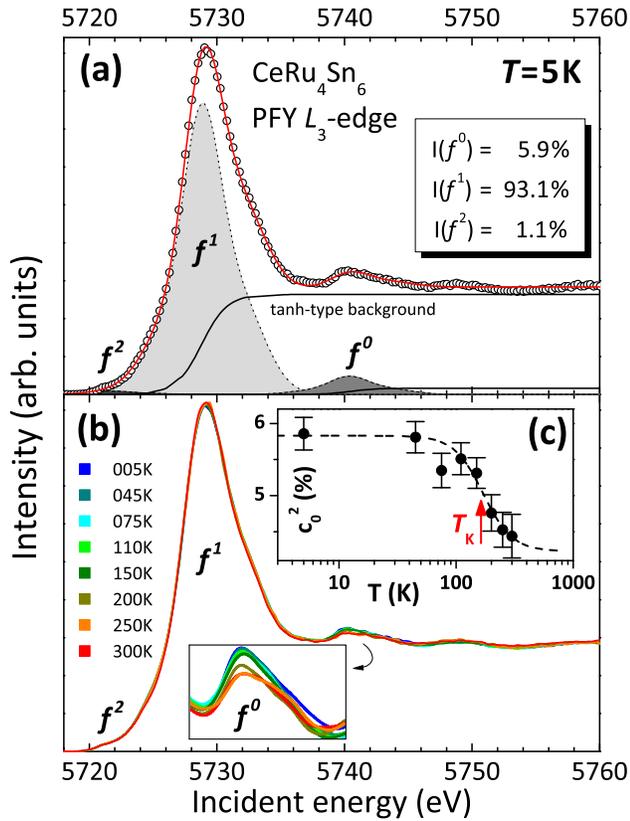}
	\caption{(a) PFY-XAS spectra of the Ce $L$-edge at $T$\,=\,5\,K (open circles). The size of the circles indicates the error bars. The red line is the total fit, the regions of different gray refer to the $I$($f^2$), $I$($f^1$) and $I$($f^0$) spectral weights, taking into account a tanh-type background (black solid lines). (b) Temperature evolution of the Ce $L_3$-edge spectra and the $I$($f^0$) peak region on an enlarged scale. (c) Inset with $I$($f^0$)\,$\simeq$\,$c_0^2$ as function of log($T$). The relative error bars amount to $\pm$0.25\%. The black, dashed line is a guide to the eye, the red arrow indicates its inflection point which is taken as Kondo temperature $T_K$.}
	\label{Fig01}
\end{SCfigure}

Figure~\ref{Fig01}(a) shows the PFY-XAS data of CeRu$_4$Sn$_6$ at 5\,K (open circles). The strongest absorption is due to the $f^1$ contribution in the ground state, the small but well resolved hump at about 5740\,eV is due to the $f^0$ contribution, and the small shoulder on the low energy tail of the $f^1$ peak (just above 5720\,eV) is due to some $f^2$ spectral weight. It is not straightforward to describe the $L_3$-edge spectral line shape since it is determined by the unoccupied $5d$ density of states in the presence of a $2p$ core hole. Yet, for extracting the $f^n$ weights, it is sufficient to fit the data with a set of Gaussian or similar functions as a mean to obtain their integrated intensities\cite{RueffPRL106,YamaokaPRL107,Kotani_2014}. We have applied such a fit by describing each spectral weight ($I$($f^0$), $I$($f^1$), and $I$($f^2$)) with identical line shapes, consisting of three Gaussian profiles each. Only the total intensities and central energy positions were varied. Each profile includes a tanh-type step function (black lines) with the maximum slope at the peak maximum to fit the background. The red line in Fig.~\ref{Fig01}(a) is the result of such a fit.

For the spectral weights we obtain $I$$(f^0)\!=\!0.059 \pm 0.015$, $I$$(f^1)\!=\!0.931 \pm 0.015$ and $I$$(f^2)\!=\!0.011 \pm 0.015$ which corresponds to $n_f \simeq 0.953$ and $c_0^2 \simeq 0.059 \pm 0.015$ under the assumption I$(f^n) \simeq c_n^2$. This is an approximation since it neglects hybridization effects in the final states\cite{Kotani_2014}. As in core level  photoemission\cite{Strigari_2015}, $I$($f^2$) overestimates $c_2^2$ and $I$($f^0$) underestimates $c_0^2$ so that the above values for $n_f$ and $c_0^2$ should be taken as upper and lower limits, respectively. It is therefore safe to conclude that the $cf$-hybridization is very strong in CeRu$_4$Sn$_6$.

Figure~\ref{Fig01}(b) shows the temperature dependence of the PFY-XAS data, normalized to the integrated intensity. The inset in Fig.~\ref{Fig01}(b) shows the $f^0$ region on an enlarged scale after smoothing the data, and it is clearly seen that the main changes occur within the temperature interval of 150 to 250\,K. Applying the same fit by varying only the total intensity of the $f^1$ and $f^0$ contributions and adjusting the height of the background according to the $f^1$ absorption line, yields the temperature variation $c_0^2(T)$ [upper inset, Fig.~\ref{Fig01}(c)]. The dashed line is a broadened step function describing the temperature evolution. We determine $T_K$ empirically as the inflection point of $c_0^2(T)$\cite{Bickers_1987,Tjeng_1993,Dallera_2002}, yielding $T_K$\,=\,$(170 \pm 25)$\,K supporting the analysis of the temperature dependence of the single crystal resistivity in Ref.\,\citenum{Guritanu_2013}.

\begin{SCfigure}
\centering
\includegraphics[width=0.6\linewidth]{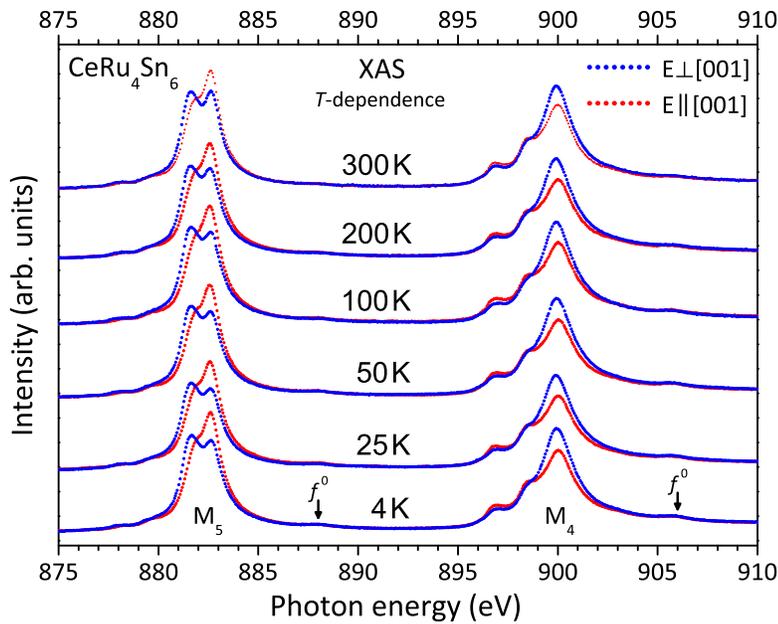}
	\caption{Temperature dependence of the linearly polarized $M$-edge XAS spectra of CeRu$_4$Sn$_6$. The arrows mark the spectral weight due to the $f^0$ contribution in the ground state.}
	\label{Fig02}
\end{SCfigure}

\begin{SCfigure}
	\centering
	\includegraphics[width=0.5\linewidth]{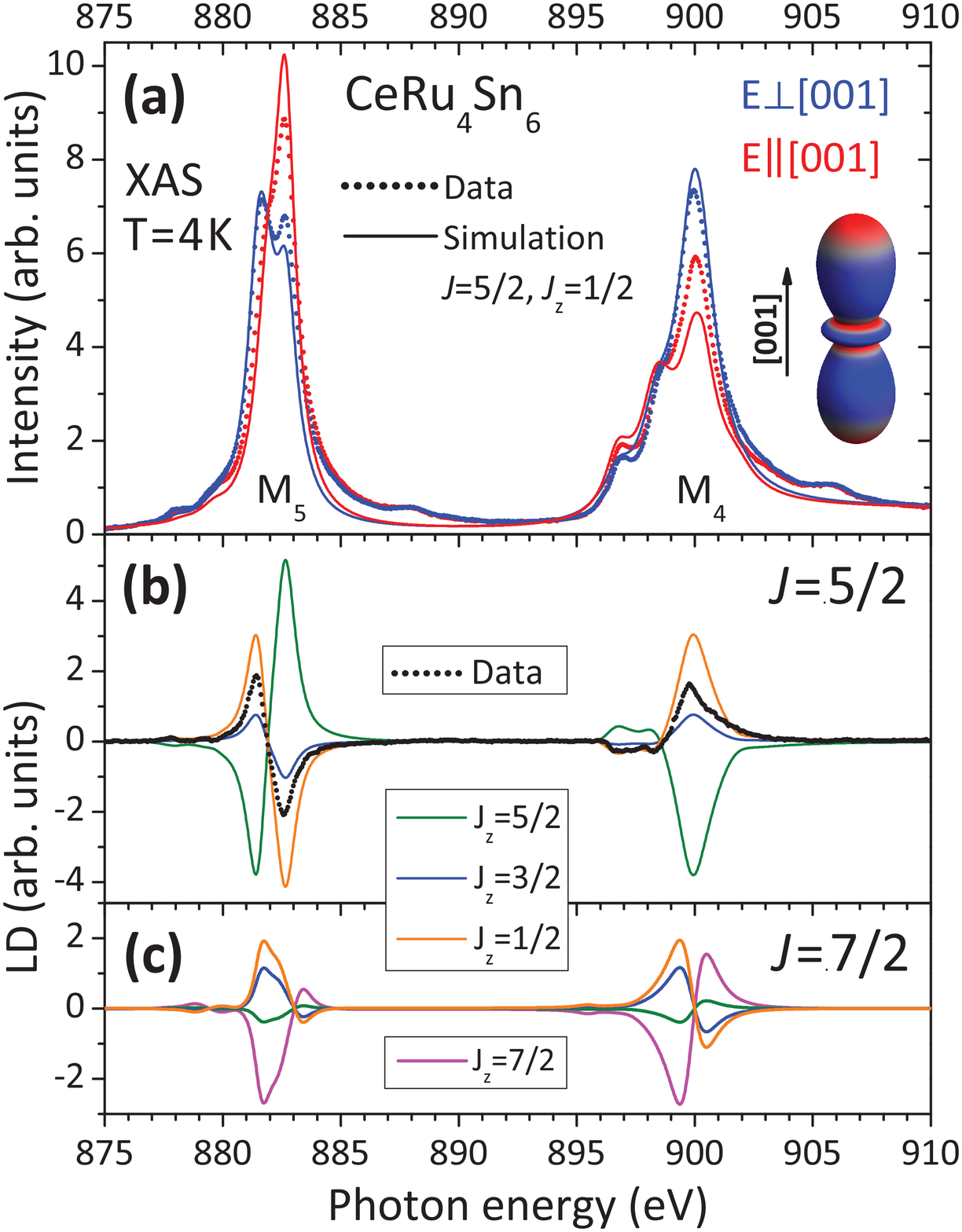}
	\caption{(a) Linearly polarized XAS data (dots) at the Ce $M$-edges for $T$\,=\,4\,K and simulation with a $\Gamma_6$\,=\,$|5/2,\pm1/2\rangle$ ground state (lines). Blue corresponds to the polarization $E$\,$\parallel$\,[001] and red to $E$\,$\perp$[001]. (b) Experimental (dots) and simulated (colored lines) linear dichroism (LD), I$_{E\perp[001]}-\mathrm{I}_{E\parallel[001]}$, of the pure $J_z$ states of the $J$\,=\,5/2 multiplet. (c) Same as (b) for the $J$\,=\,7/2 multiplet. The inset in (a) shows the spatial distribution of the $4f$ electrons for $\Gamma_6$ symmetry.}
	\label{Fig03}
\end{SCfigure}

\subsection*{CEF ground state symmetry}
The local point symmetry at the Ce site of $D_{2d}$\cite{Pottgen_1997} in CeRu$_4$Sn$_6$ gives rise to a tetragonal CEF that splits the $J$\,=\,5/2 Hund's rule ground state into three Kramers' doublets: one of the type $\Gamma_6\!=\!|5/2,\pm 1/2\rangle$, and two mixed states of type $\Gamma_7^1\!=\!\sqrt{1-\alpha^2}$\,$|5/2,\pm3/2\rangle+\alpha$\,$|5/2,\mp5/2\rangle$ and $\Gamma_7^2\!=\!\sqrt{1-\alpha^2}$\,$|5/2,\pm5/2\rangle-\alpha$\,$|5/2,\mp3/2\rangle$, in the $J_z$-representation $|J,J_z\rangle$. We have investigated the CEF ground state wave function with two spectroscopic methods: soft XAS at the Ce $M$-edge ($3d \rightarrow 4f$)\cite{HansmannPRL100,WillersPRB80,Strigari_2012} and hard x-ray NIXS at the Ce N-edge ($4d \rightarrow 4f$)\cite{WillersPRL109}. In the soft XAS experiment, the dipole selection rules for linearly polarized light give access to the initial state symmetry. Excited CEF states only contribute to the polarization as they get increasingly populated with rising temperature. A NIXS experiment with hard x-rays is truly bulk sensitive, and when performed at large momentum transfers $|\vec{q}|$ the signal is dominated by higher multipole orders of the scattering cross section\cite{Gordon2008,GordonJElecSpec184}. In the NIXS experiment the $\vec{q}$-dependence ($\vec{q}$-direction with respect to the sample orientation) replaces the polarization dependence in XAS and, accordingly, $\vec{q}$-dependent multipole selection rules give access to the ground state symmetry\cite{WillersPRL109}.

Figure~\ref{Fig02} shows the soft XAS data at the Ce M$_{4,5}$ edge for the two polarizations $E$\,$\parallel$\,[001] and $E$\,$\perp$[001] for several temperatures between 4 and 300\,K. At all temperatures there is a strong polarization effect which decreases only slightly between 200 and 300\,K, thus indicating the first excited CEF state is situated at $\geq$\,25\,meV. The data at 4\,K will therefore yield ground state information. In addition to the main absorption line, there are smaller humps at slightly higher photon energies (see arrows) due to some $f^0$ weight in the initial state. Its spectral weight decreases with rising temperature in agreement with the findings from the PFY-XAS data.

Figure~\ref{Fig03}(a) shows the linearly polarized low-temperature data (red and blue dots). For the simulation of the spectra a full multiplet code is used (see \textsl{Methods, Analysis}). Figure~\ref{Fig03}(a) shows a simulation assuming a pure $|5/2,\pm1/2\rangle$\,=\,$\Gamma_6$ ground state (red and blue lines), which is the state with rotational symmetry [see inset in Fig.~\ref{Fig03}(a)]. The simulation reproduces the main features of the spectra and their polarization dependence, but the match is by no means perfect. This issue is investigated further in the following. 

In Fig.~\ref{Fig03}(b) the experimental linear dichroism (LD) ($I$$_{E\perp[001]}$-$I$$_{E\|001}$, black dots) is compared with the simulated LD of the three pure $|5/2,J_z\rangle$ states (colored lines). The LD of the pure $|5/2,\pm1/2\rangle$\,=\,$\Gamma_6$ (orange) and also of the $|5/2,\pm3/2\rangle$ (blue), corresponding to a $\Gamma_7$-symmetry with $\alpha$\,=\,0, have the correct sign, but the LD of the latter is too small and any $|5/2,\mp5/2\rangle$-admixture (green) in a $\Gamma_7$-type ground state would make the discrepancy even larger. Even if we adopt a $\Gamma_7$-type ground state of predominantly $|5/2,\pm3/2\rangle$, the measured LD will be larger than the simulated one. We therefore conclude that although the ground state is not a pure $\Gamma_6$\,=\,$|5/2,\pm1/2\rangle$, it is still mainly of that character.

\begin{SCfigure}
 \centering
 \includegraphics[width=0.5\linewidth]{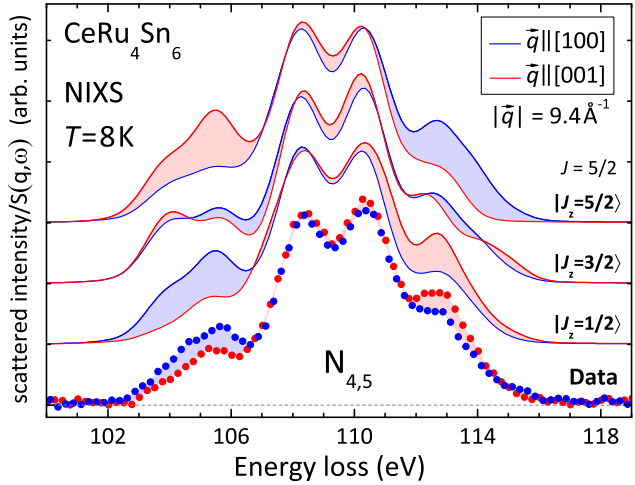}
	\caption{Dots: Experimental NIXS intensity at the Ce $N_{4,5}$ edge after subtracting a linear background. The point size reflects the statistical error, the counting time was 240\,s per point. Lines: Calculated scattering function S($\vec{q},\omega$) for the pure $J_z$ states of the $J$\,=\,5/2 multiplet. Data and simulation are shown for $\vec{q}$\,$\parallel$\,[100] (blue) and $\vec{q}$\,$\parallel$\,[001] (red).}
	\label{Fig04}
\end{SCfigure}

A NIXS experiment at the Ce N-edge was performed in order to rule out surface effects being responsible for the unsatisfactory agreement between the soft XAS data and the simulation. The data in Fig.~\ref{Fig04} (red and blue dots) were obtained after 240\,s counting per point and after subtracting a rising, linear background (Compton). The colors are chosen as before, i.e$.$ red for $\vec{q}$\,$\parallel$\,[001] and blue for $\vec{q}$\,$\parallel$\,[100]. Again there is a clear dichroic effect. We compare the measured $N_{4,5}$ edge with simulations for the pure $|5/2,J_z\rangle$ states (red and blue lines), which here are shown again (cf$.$ Ref.\,\citenum{WillersPRL109}) for reasons of convenience. Details about the theoretical simulations containing a full multiplet ansatz can be found in Section \textsl{Methods, Analysis}.

The strongest dichroism in the data appears at the shoulder structures at about 105 and 113\,eV. Qualitatively the simulation for the $|5/2,\pm1/2\rangle$ state reproduces the data quite well. In contrast, the $|5/2,\pm3/2\rangle$ state suggests a crossing of both directions on both shoulders, which is not observed experimentally, and the dichroism of the $|5/2,\pm5/2\rangle$ state has the wrong sign. We therefore conclude also from this experiment that the ground state must be governed by a $\Gamma_6$\,=\,$|5/2,\pm1/2\rangle$ symmetry, but at the same time is not purely given by the $\Gamma_6$.

\section*{Discussion}
What could be the reason for the deviation between the experiment and the simulation starting from a pure $\Gamma_6$\,=\;$|5/2,\pm1/2\rangle$? Already in 1986 van der Laan \textsl{et al.} pointed out that in the presence of strong hybridization higher lying states contribute to the $M$-edge signal\,\cite{Laan_1986}. In CeRu$_4$Sn$_6$ we find that the Kondo temperature (170 K) is not much smaller than the CEF excitations ($\geq$\,25\,meV\,$\hat{=}$\,290\,K) so that a contribution of higher lying CEF states to the ground state is indeed to be expected. Given the fact that the LD of these higher lying CEF states is smaller or even of opposite sign, see Fig.~\ref{Fig03}(b) and (c), it is evident that the LD of this composite ground state will be reduced from that of the pure $|5/2,\pm1/2\rangle$  state. We could also reverse the argumentation: the observation of a reduced LD can be taken as evidence for the presence of strong hybridization. Nevertheless, we can conclude that the ground state of CeRu$_4$Sn$_6$ is predominantly of $|5/2,\pm1/2\rangle$ symmetry. This is in good agreement with the anisotropy of the low-temperature magnetization\,\cite{Winkler_2012} and susceptibility \cite{Sidorenko-unp}, thereby supporting  also the results of LDA+DMFT calculations at 290\,K\,\cite{Wissgott_Cond_Mat}.

\begin{figure}
	\centering
	\includegraphics[width=0.52\linewidth]{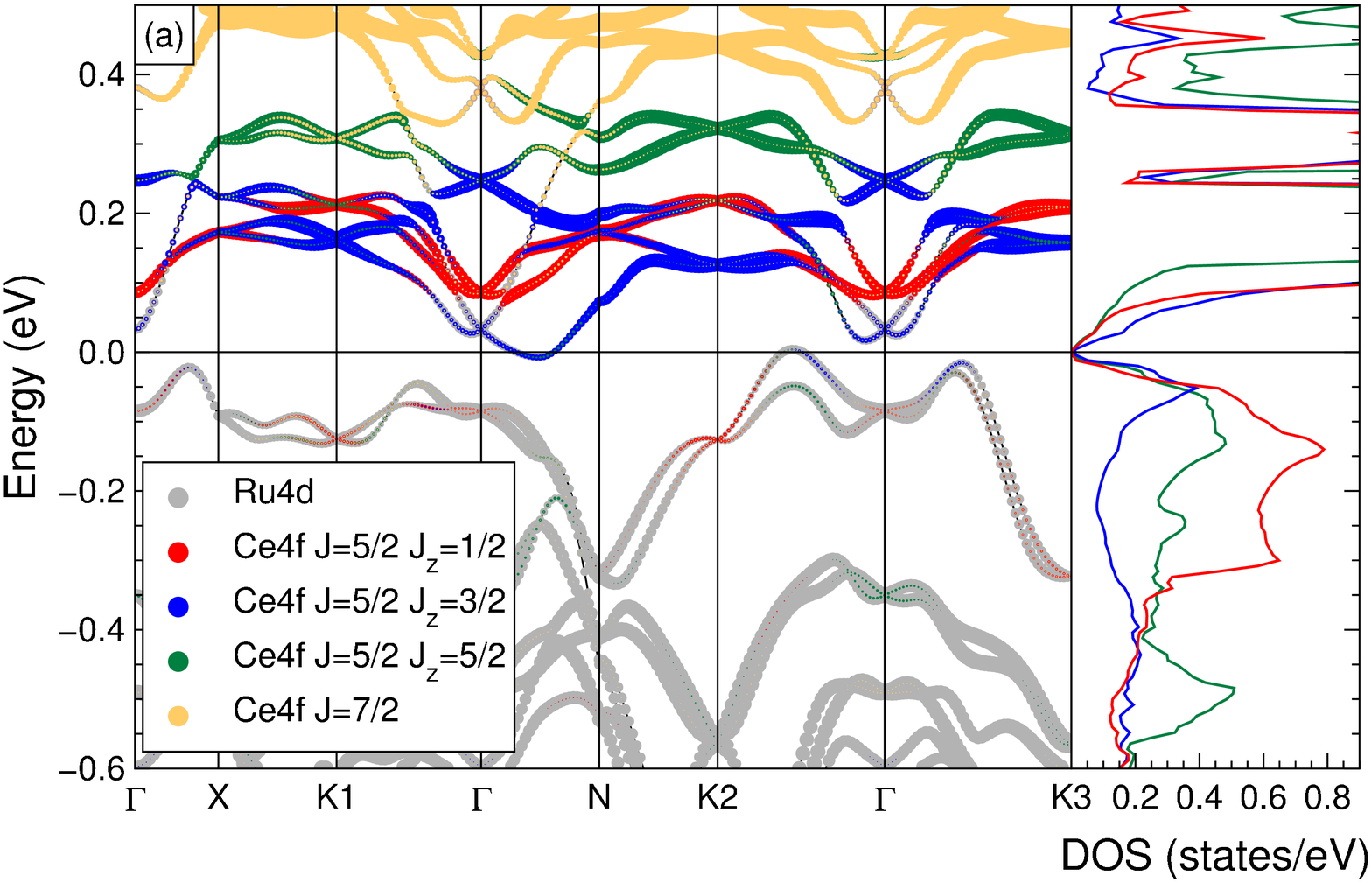}
	\centering
	\includegraphics[width=0.495\linewidth]{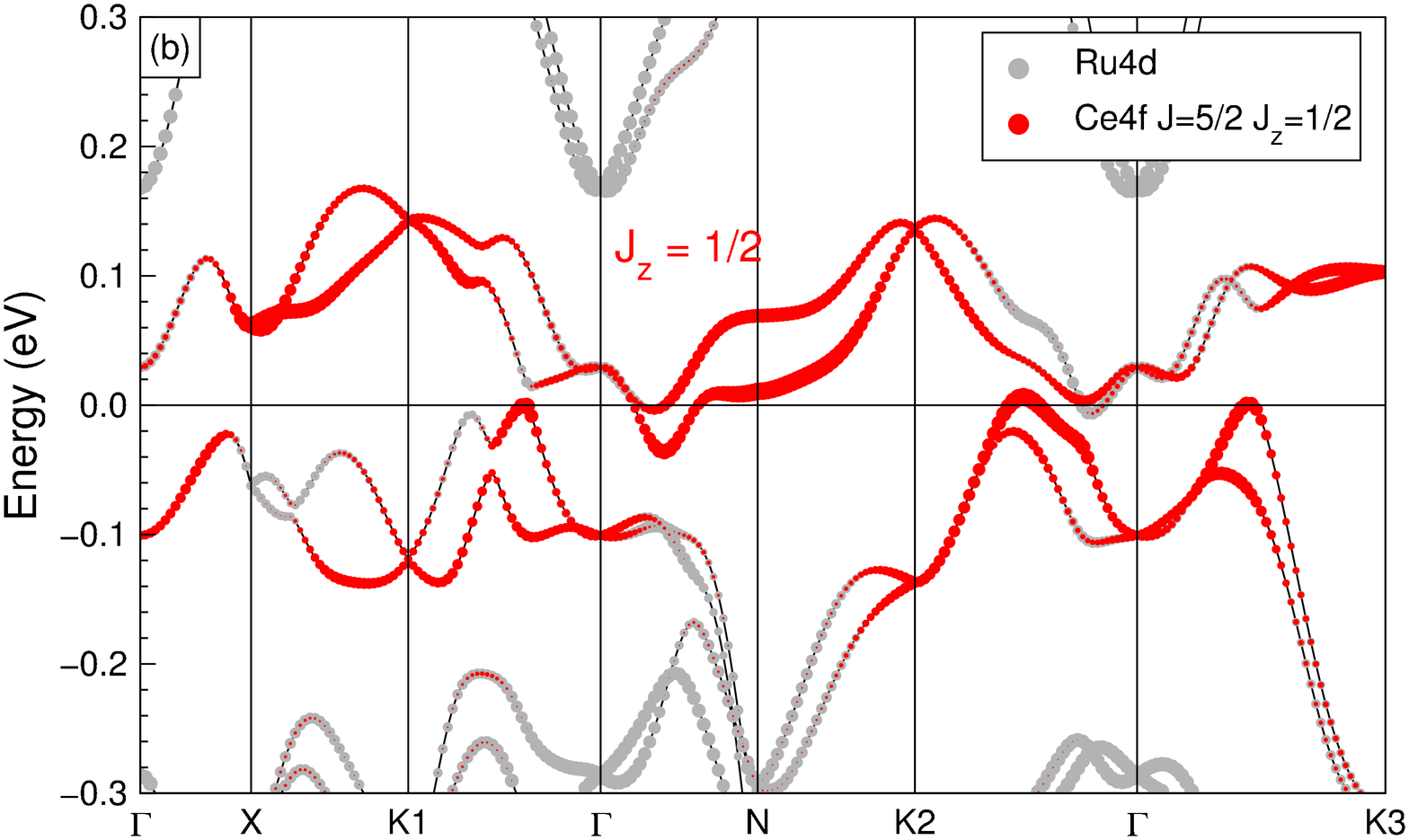}
	\centering
	\includegraphics[width=0.495\linewidth]{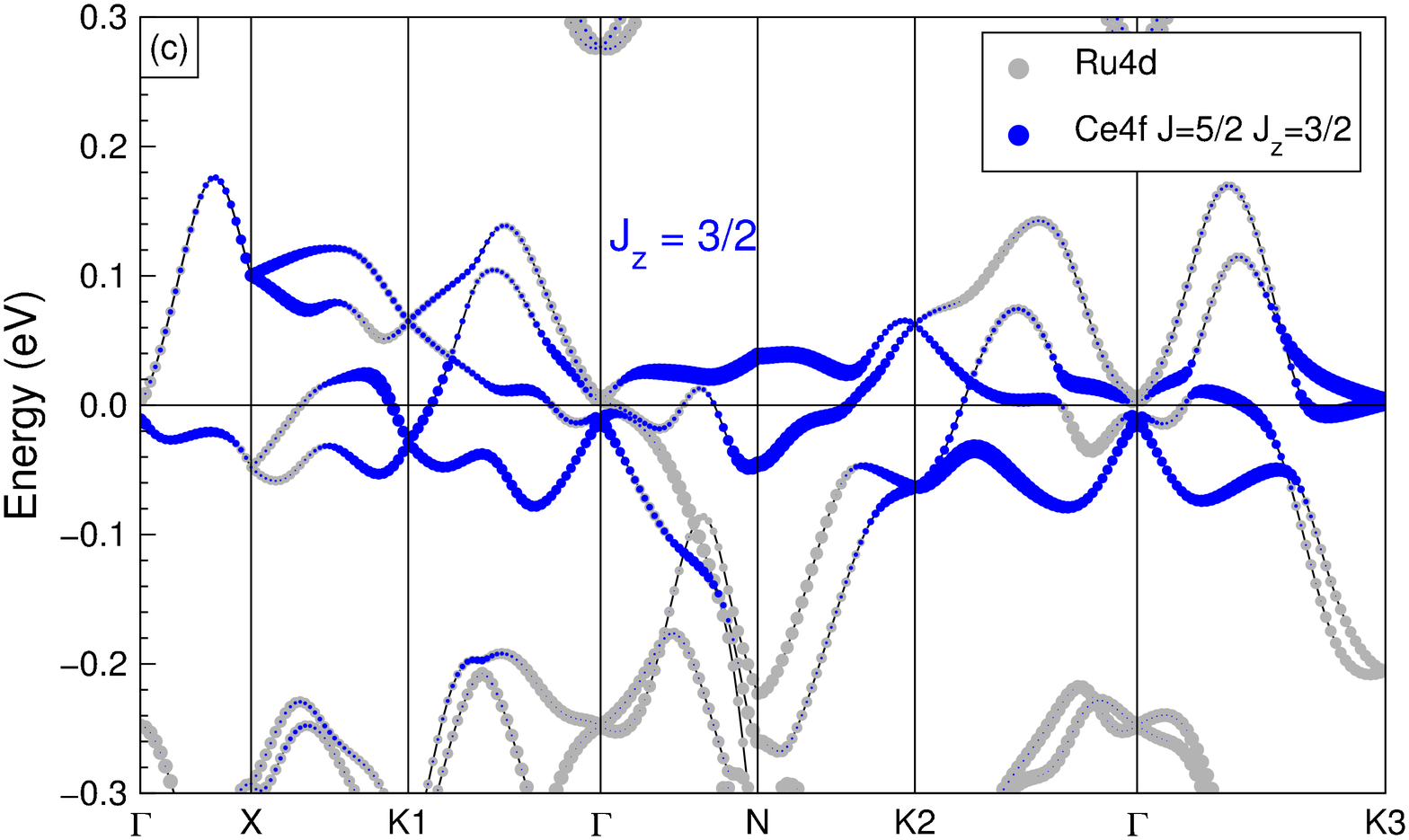}
	\caption{(a) Full relativistic (LDA+spin-orbit) band structure of CeRu$_4$Sn$_6$, wherein the Ce 4$f$ weight is mostly above the Fermi level. From the $J_{z}$ projected DOS, we note the largest contribution to the occupied valence sector is from $J$ = 5/2 and $J_{z}$ = $\pm$ 1/2, (b) Applying a constant potential to the $J_z$\,=\,1/2 state. (c) Applying a constant potential to the $J_z$\,=\,3/2 state.}
	\label{Fig05}
\end{figure}

We now will use density functional theory based band structure calculations to obtain a deeper understanding of the experimental findings (details in Section \textsl{Materials and Methods, Band structure calculations}). In Fig.~\ref{Fig05}\,(a) we display our result of a full-relativistic non-magnetic calculation employing the local density approximation (LDA) for CeRu$_4$Sn$_6$. We can clearly observe that the band structure has a true band gap in agreement with an earlier study in \,\cite{Guritanu_2013}, comprising of unoccupied Ce $4f$ bands above the Fermi level and occupied Ru $4d$ bands below the Fermi level. The presence of the gap is important since it confirms that the system has the even number of electrons per unit cell necessary to allow for the formation of a fully spin-compensated, non-magnetic state. We also notice in the left panel of Fig.~\ref{Fig05}\,(a) that the lowest unoccupied Ce $4f$ state is surprisingly given by the $J_z$\,=\,$|5/2,\pm3/2\rangle$ (blue dots) and not by the $|5/2,\pm1/2\rangle$ (red dots). Yet, looking at the Ce $4f$ partial density of states depicted in the right panel of Fig.~\ref{Fig05}\,(a) we find that it is the $|5/2,\pm1/2\rangle$ which has the largest occupation, as found in our experiment. We therefore can infer from these calculations that the empty $|5/2,\pm1/2\rangle$ has indeed the stronger hybridization with the occupied Ru $4d$ bands so that it gets pushed up higher in energy and, at the same time, acquire a larger occupation.

Next, we will make an attempt to estimate the effect of correlations on the relevant symmetries of the system. We do this by introducing artificially an extra potential for a particular $J_z$ state so that it becomes more occupied, i.e. as to mimic the Ce $4f^1$ valence state thereby making the inversion with the Ru $4d$. Figure~\ref{Fig05}\,(b) shows the result for moving down the $|5/2,\pm1/2\rangle$ orbital and Fig.~\ref{Fig05}\,(c) the $J_z$\,=\,$|5/2,\pm3/2\rangle$. We can observe that in the $J_z$\,=\,$|5/2,\pm3/2\rangle$ case the system becomes metallic with quite a large number of bands crossing the Fermi level\,\cite{footnote1}. For the the $|5/2,\pm1/2\rangle$ case, we also see that the system is metallic, but only with a few crossings and small Fermi surface areas, suggesting that the number of charge carriers will be low. Important for the topology is that the band structure is that of a semi-metal, i.e. that there is a warped gap, so that a unique assignment can be made about the character of the bands. This aspect did not necessarily follow from the conditions set by Dzero and Galitzki \cite{Dzero_2012,Dzero_2013}, since there could always have been some other non-bonding bands crossing the Fermi level, which is not considered in their effective model. Yet, our calculations indicate specifically that such is not the case for CeRu$_4$Sn$_6$. It is also important to notice that the warped gap is substantial, i.e. 20 meV or larger, which corroborates nicely with the fact that the $J_z$\,=\,$\pm 1/2$ is nodeless and therefore guarantees the presence of hybridization with the Ru $4d$ bands at all $k$-points of the Brillouin Zone (except at the time reversal inversion points).

To summarize, we thus have found that CeRu$_4$Sn$_6$ fulfills the conditions mentioned by Dzero and Galitzki\cite{Dzero_2012,Dzero_2013} to be a strong topological insulator. It has a ground state symmetry given primarily by the $\Gamma_6$\,=\,$|5/2,\pm1/2\rangle$ and has a $4f$ occupation sufficiently far away from integer with a very high Kondo temperature (170\,K) so that the effective degeneracy is larger than two. We have in addition verified that there are no non-bonding bands that otherwise could spoil the formation of a warped gap. We therefore conclude that CeRu$_4$Sn$_6$ is a strongly correlated material with non-trivial topology.

\section*{Methods}
\subsection*{Samples} All experiments were performed on single crystals. Single crystals were grown with the self-flux floating-zone melting method using optical heating in a four-mirror furnace\cite{Prokofiev_2012}. The quality of the crystals was checked with Laue and x-ray diffraction as well as scanning electron microscopy and energy dispersive x-ray spectroscopy, which show that the samples are single grain and single phase materials. The lattice parameters are in agreement with $a$\,=\,6.8810\,\AA\ and $c$\,=\,9.7520\,\AA\ as given in Ref.\,\citenum{Pottgen_1997}. The single crystals were pre-aligned with a Laue camera. A peculiarity of the present tetragonal structure is that $\sqrt{2}$$a$\,$\approx$\,$c$ within $\approx$\,0.2\%. As a consequence the structure appears pseudo-cubic in Laue images and thus the final alignment of the single crystals had to be done via magnetization measurements\cite{Paschen_2010}.

\subsection*{Set-ups of x-ray techniques} The PFY-XAS experiment was performed at the GALAXIES inelastic scattering endstation at the SOLEIL synchrotron in France\cite{GalaxiesPaper}. The syn\-chro\-tron radiation was monochromatized using a Si(111) nitrogen-cooled fixed-exit double-crystal monochromator ($\Delta E/E\!\approx\!1.4 \times 10^{-4}$), followed by a Pd-coated spherical collimating mirror. The x-rays were then focused to a spot size of 30\,$\mu$m\,x\,90\,$\mu$m FWHM [vertical\,x\,horizontal] at the sample position by a 3:1 toroid Pd-coated mirror. The overall energy resolution for the experiment was measured to be 700\,meV FWHM with about 1$\times$10$^{13}$ photons/second on the sample. The incident x-ray energy was scanned across the Ce L$_3$ absorption edge ($\approx$\,5723\,eV) while monitoring the intensity of the Ce L$_{\alpha_1}$ de-excitation ($\approx$\,4839\,eV) using a Ge\,331 spherical bent crystal analyzer (Bragg angle 80.8$^{\circ}$, 1\,m radius of curvature) in the Rowland circle geometry which focused the emitted x-rays onto a silicon avalanche photodiode. All measurements took place in a backscattering geometry and normalization was performed using a monitor placed just before the sample that recorded the scattered radiation from a Kapton foil placed in the beam path. The flight path was filled with helium gas to reduce the air scattering and the samples were cooled in a helium cryostat.

The XAS spectra were recorded at the synchrotron light source BESSY~II using the UE46 PGM-1 undulator beamline. The total electron yield (TEY) was measured under ultra high vacuum of the order of 10$^{-10}$\,mbar. The CeRu$_4$Sn$_6$ single crystals were cleaved \textsl{in situ} to guarantee clean sample surfaces. The TEY signal was normalized to the incoming photon flux $I_0$ as measured at the refocusing mirror. The energy resolution at the cerium M$_{4,5}$ edges was set to 0.15\,eV. The undulator combined with a normal incident measurement geometry allow a change in polarization without changing the probing spot on the sample surface. The crystals were oriented such that the two polarizations $E$\,$\parallel$\,[001] and $E$\,$\perp$\,[001], [001] being the long tetragonal axis, could be measured. Several crystals were measured or re-cleaved in order to assure the reproducibility of the data.

The NIXS measurements were performed at the beamline ID20 at ESRF. A single Si(111) monochromator set the incident energy to 9690\,eV and the scattered intensity was analyzed by three columns of three Si(660) analyzers (in total nine) at \textsl{in-plane} scattering angles of $2\,\Theta = 140^\circ, 146^\circ$, and $153^\circ$, corresponding to an averaged momentum transfer of $|\vec{q}|\!\approx\!(9.4 \pm 0.2)$\,\AA$^{-1}$. This configuration yields an instrumental energy resolution of FWHM\,=\,1.3\,eV and made the Ce N$_{4,5}$ edge measurable beyond the dipole limit\cite{Gordon2008,GordonJElecSpec184,WillersPRL109}. Two crystals with polished (100) and (001) surfaces were mounted in a cryostat and measured in specular geometry with $\vec{q}$\,$\|$\,[001] and $\vec{q}$\,$\bot$\,[100], i.e$.$ parallel and perpendicular to the $c$ axis.

\subsection*{Analysis} 
\begin{SCfigure}
\centering
\includegraphics[width=0.5\linewidth]{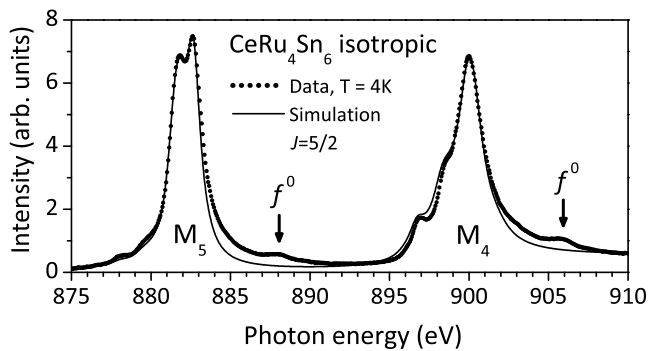}
	\caption{Dots: isotropic spectrum as constructed from the experimental data $I_{iso}$\,=\,1/3\,($I_{E\parallel[001]}$\,+\,2$I_{E\perp[001]}$). Solid line: fit with full muliplet routine after adjusting reduction factors (see text).}
	\label{Fig06}
\end{SCfigure}

The soft XAS $M$-edge data were analyzed with the full multiplet code \textsc{XTLS}~8.3 as described in Ref.\,\citenum{HansmannPRL100}. The atomic parameters are given by Hartree-Fock values and the reduction factors for the Coulomb and spin orbit interaction interaction are determined by fitting the isotropic spectrum I$_\mathrm{iso}$\,=\,1/3\,(I$_{E\parallel[001]}$\,+\,2I$_{E\perp[001]}$) in Fig.~\ref{Fig06}. We obtain reduction factors of 41\,\% for the $4f$-$4f$ and 20\,\% for the $3d$-$4f$ Coulomb interaction while the spin orbit values remain unchanged within 1 or 2\%. These are typical values for Ce intermetallic compounds\cite{HansmannPRL100,WillersPRB80,Strigari_2012}. For the reproduction of the polarized data, the CEF has to be taken into account. The calculations are purely ionic.

The NIXS data were simulated using the full multiplet code \textsl{Quanty}\cite{MauritsQuanty}. A Gaussian and a Lorentzian broadening of FWHM\,=\,1.3\,eV and 0.2\,eV, respectively, are assumed to account for the instrumental resolution and life time effects. The atomic Hartree-Fock values were adjusted via the peak positions, resulting in reductions of 30\,\% and 20\,\% for the 4$f$-4$f$ and 4$d$-4$f$ Coulomb interactions, respectively. The simulation procedure is explained in detail in Ref.\,\citenum{WillersPRL109}.

\subsection*{Band structure calculations}
Density functional theory based calculations were performed using the full-potential non-orthogonal local orbital (\texttt{FPLO})\cite{Koepernik_1999} code employing the local density approximation (LDA) and including spin-orbit coupling. The band structure and density of states (DOS) are obtained from a carefully converged calculation with 1000 $k$-points in the irreducible wedge of the Brillouin zone. For easy comparison with literature, we have plotted the band structure along the $k$-path as described in Ref.\,\citenum{Wissgott_Cond_Mat}.
To simulate the enhancement in the occupation of the various $J_{z}$ states, we have applied a constant potential shift to the on-site energies (1 eV) of the relevant orbitals.

\section*{Acknowledgment}
We thank K.~Held, P.~Thalmeier and D.T.~Adroja for fruitful discussions. M.S., F.S, T.W. and A.S. benefited from support of the German funding agency DFG (Projects 600575 and 615811) and H.W., A.P. and S.P. from the Austrian Science Fund (Project I623-N16) and the U.S. Army Research Office (Grant W911NF-14-1-0497). We further acknowledge SOLEIL for provision of synchrotron radiation facilities (proposal 20120919).

\section*{Author contributions statement}
A.S. and L.H.T designed the research, 
M.S., F.S., T.W., J.M.A., J.-P.R., D.S., E.W, M.M.S., A.A-Z., L.H.T, and A.S. performed the experiments, 
H.W., A.P., and S.P. made high quality singly crystals, 
A.T. and M.W.H. contributed codes for data analysis, 
M.S., F.S., D.K., and A.S. analyzed data, 
M.S., L.H.T., S.P. and A.S. wrote the paper.
All authors reviewed the manuscript.



\end{document}